\documentclass[twocolumn,aps,prc,showpacs,floatfix]{revtex4}
\usepackage{graphicx}
\usepackage{dcolumn}
\usepackage{bm}

\begin{document}

\title{Nuclear symmetry energy and proton-rich reactions at intermediate energies}
\author{Gao-Chan Yong}
\affiliation{Institute of Modern Physics, Chinese Academy of
Sciences, Lanzhou 730000, China}

\begin{abstract}
Based on an isospin dependent transport model IBUU, effects of
high density behavior of nuclear symmetry energy in the
proton-rich reaction $^{22}$Si+$^{22}$Si at a beam energy of $400$
MeV/nucleon are studied. It is found that the symmetry energy
affects $\pi^{+}$ production more than $\pi^{-}$. More
interestingly, comparing with neutron-rich reactions, for
$\pi^{-}/\pi^{+}$ and dense matter's N/Z ratios, effects of
symmetry energy in the proton-rich reaction both show contrary
behaviors. The practical experiment by using the proton-rich
reaction $^{22}$Si+$^{40}$Ca to study nuclear symmetry energy are
also provided.
\end{abstract}

 \maketitle


Recently, pion production in heavy-ion collisions has attracted
much attention in nuclear physics community
\cite{xiao09,ditoro1,xu09,Rei07,yong06}. One important reason is
that pion production is connected with the high density behavior
of nuclear symmetry energy \cite{LiBA02}. The latter is crucial
for understanding many interesting issues in both nuclear physics
and astrophysics
\cite{Bro00,Dan02a,Bar05,LCK08,Sum94,Lat04,Ste05a}. The high
density behavior of nuclear symmetry energy, however, has been
regarded as the most uncertain property of dense neutron-rich
nuclear matter \cite{Kut94,Kub99}. Many microscopic and/or
phenomenological many-body theories using various interactions
\cite{Che07,LiZH06} predict that the symmetry energy increases
continuously at all densities. On the other hand, other models
\cite{Pan72,Fri81,Wir88a,Kra06,Szm06,Bro00,Cha97,Sto03,Che05b,Dec80,MS,Kho96,Bas07,Ban00}
predict that the symmetry energy first increases to a maximum and
then may start decreasing at certain supra-saturation densities.
Thus, currently the theoretical predictions on the symmetry energy
at supra-saturation densities are extremely diverse. To make
further progress in determining the symmetry energy at
supra-saturation densities, what is most critically needed is some
guidance from dialogues between experiments and transport models,
which have been done extensively in the studies of nuclear
symmetry energy at low densities
\cite{tsang09,shetty07,fami06,tsang04,chen05}.

Using $\pi^-/\pi^+$ to probe the high density behavior of nuclear
symmetry energy has evident advantage within both the $\Delta$
resonance model and the statistical model \cite{Sto86,Ber80}. And
several hadronic transport models have quantitatively shown that
$\pi^-/\pi^+$ ratio is indeed sensitive to the symmetry energy
\cite{LiBA02,yong06,Gai04,LiQF05b}, especially around pion
production threshold. These transport models, however, usually
simulate neutron-rich reactions to study the effect of symmetry
energy, few proton-rich reactions on this subject were reported.
Here we study the effects of symmetry energy on pion production,
as well as the value of $\pi^{-}/\pi^{+}$ in the proton-rich
collision because the National Superconducting Cyclotron
Laboratory at Michigan State University, Rikagaku Kenkyusho
(RIKEN, The Institute of Physical and Chemical Research) of Japan,
and the Cooler Storage Ring in Lanzhou, China, are planning to do
experiments of pion production to study the high density behavior
of nuclear symmetry energy. In the framework of an
Isospin-dependent Boltzmann-Uehling- Uhlenbeck (IBUU) transport
model, as an example, we studied the effects of symmetry energy on
$\pi^{-}/\pi ^{+}$ in the proton-rich reaction $^{22}$Si+$^{22}$Si
at a beam energy of $400$ MeV/nucleon. It is found that the
symmetry energy affects the value of $\pi^{-}/\pi^{+}$ and dense
matter's N/Z ratios. As one expected, comparing with neutron-rich
reactions, effect of symmetry energy in the proton-rich reaction
shows contrary behaviors.


The isospin and momentum-dependent mean field potential used in
the present work is \cite{Das03,IBUU04}
\begin{eqnarray}
U(\rho, \delta, \textbf{p},\tau)
=A_u(x)\frac{\rho_{\tau^\prime}}{\rho_0}+A_l(x)\frac{\rho_{\tau}}{\rho_0}\nonumber\\
+B\left(\frac{\rho}{\rho_0}\right)^\sigma\left(1-x\delta^2\right)\nonumber
-8x\tau\frac{B}{\sigma+1}\frac{\rho^{\sigma-1}}{\rho_0^\sigma}\delta\rho_{\tau^{\prime}}\nonumber\\
+\sum_{t=\tau,\tau^{\prime}}\frac{2C_{\tau,t}}{\rho_0}\int{d^3\textbf{p}^{\prime}\frac{f_{t}(\textbf{r},
\textbf{p}^{\prime})}{1+\left(\textbf{p}-
\textbf{p}^{\prime}\right)^2/\Lambda^2}},
\label{Un}
\end{eqnarray}
where $\rho_n$ and $\rho_p$ denote neutron ($\tau=1/2$) and proton
($\tau=-1/2$) densities, respectively.
$\delta=(\rho_n-\rho_p)/(\rho_n+\rho_p)$ is the isospin asymmetry
of nuclear medium. All parameters in the above equation can be
found in refs. \cite{IBUU04}. The variable $x$ is introduced to
mimic different forms of the symmetry energy predicted by various
many-body theories without changing any property of symmetric
nuclear matter and the value of symmetry energy at normal density
$\rho_0$. In this article we let the variable $x$ be $1$. With
these choices the symmetry energy obtained from the above single
particle potential is consistent with the Hartree-Fock prediction
using the original Gogny force \cite{Das03} and is also favored by
recent studies based on FOPI experimental data \cite{xiao09}. To
study the effect of symmetry energy, we also select the stiff
symmetry energy parameter $x =0$ \cite{chen05} as reference. The
main reaction channels related to pion production and absorption
are
\begin{eqnarray}
&& NN \longrightarrow NN \nonumber\\
&& NR \longrightarrow NR \nonumber\\
&& NN \longleftrightarrow NR \nonumber\\
&& R \longleftrightarrow N\pi,
\end{eqnarray}
where $R$ denotes $\Delta $ or $N^{\ast }$ resonances. In the
present work, we use the isospin-dependent in-medium reduced $NN$
elastic scattering cross section from the scaling model according
to nucleon effective mass \cite{factor,neg,pan,gale} to study the
effect of symmetry energy on pion production. Assuming in-medium
$NN$ scattering transition matrix is the same as that in vacuum
\cite{pan}, the elastic $NN$ scattering cross section in medium
$\sigma _{NN}^{medium}$ is reduced compared with their free-space
value $\sigma _{NN}^{free}$ by a factor of
\begin{eqnarray}
R_{medium}(\rho,\delta,\textbf{p})&\equiv& \sigma
_{NN_{elastic}}^{medium}/\sigma
_{NN_{elastic}}^{free}\nonumber\\
&=&(\mu _{NN}^{\ast }/\mu _{NN})^{2}.
\end{eqnarray}
where $\mu _{NN}$ and $\mu _{NN}^{\ast }$ are the reduced masses
of the colliding nucleon pair in free space and medium,
respectively. For in-medium $NN$ inelastic scattering cross
section, even assuming in-medium $NN \rightarrow NR$ scattering
transition matrix is the same as that in vacuum, the density of
final states $D_{f}^{'}$ \cite{pan} of $NR$ is very hard to
calculate due to the fact that the resonance's potential in matter
is presently unknown. The in-medium $NN$ inelastic scattering
cross section is thus quite controversial
\cite{ditoro2,Lar01,Lar03,Ber88,Mao97}. Because the purpose of
present work is just study the effect of symmetry energy on pion
production and charged pion ratio, to simplify the question, for
the $NN$ inelastic scattering cross section we use the free $NN$
inelastic scattering cross section. The effective mass of nucleon
in isospin asymmetric nuclear matter is
\begin{equation}
\frac{m_{\tau }^{\ast }}{m_{\tau }}=\left\{ 1+\frac{m_{\tau }}{p}\frac{%
dU_{\tau }}{dp}\right\}^{-1}.
\end{equation}
From the definition and Eq.~(\ref{Un}), we can see that the
effective mass depends not only on density and asymmetry of medium
but also the momentum of nucleon.


\begin{figure}[th]
\begin{center}
\includegraphics[width=0.45\textwidth]{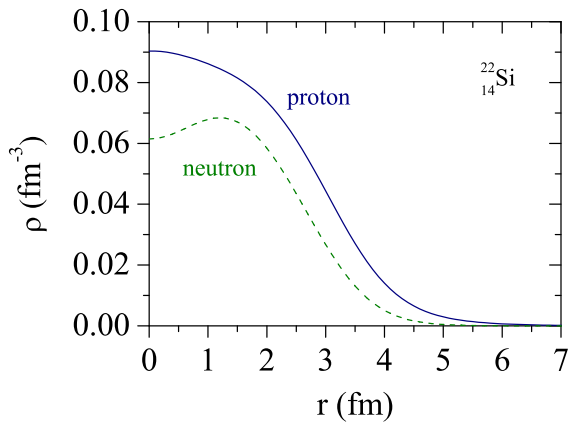}
\end{center}
\caption{(Color online) Density distributions of protons and
neutrons of the nucleus $^{22}_{14}$Si.} \label{dis}
\end{figure}
Fig.~\ref{dis} shows density distributions of protons and neutrons
of nucleus $^{22}_{14}$Si, which were given by Skyrme-Hartree-Fock
with Skyrme $M^{*}$ force parameters \cite{Friedrich86}. The
initializations of colliding nuclei does not evidently affect our
results here \cite{yong1103}. Please note here that, we do not
study the effect of nucleonic distribution on the value of
$\pi^{-}/\pi^{+}$ since the values of nuclear symmetry energy
around saturation density are still open. And besides, our study
here is just about the effects of high density symmetry energy. It
is clearly shown from Fig.~\ref{dis} that density distribution of
protons is always larger than that of neutrons. Therefore there is
no neutron skin as that in neutron-rich nucleus. We in the present
studies use two proton-rich colliding nuclei
$^{22}_{14}$Si+$^{22}_{14}$Si to produce proton-rich nuclear
matter, to study the effect of high density behavior of nuclear
symmetry energy in proton-rich nuclear matter. Such matter can
form not only in the proton-rich nuclear reaction, but also
possibly in neutron stars driven by the strong magnetic fields
\cite{Ch97}.

\begin{figure}[th]
\begin{center}
\includegraphics[width=0.45\textwidth]{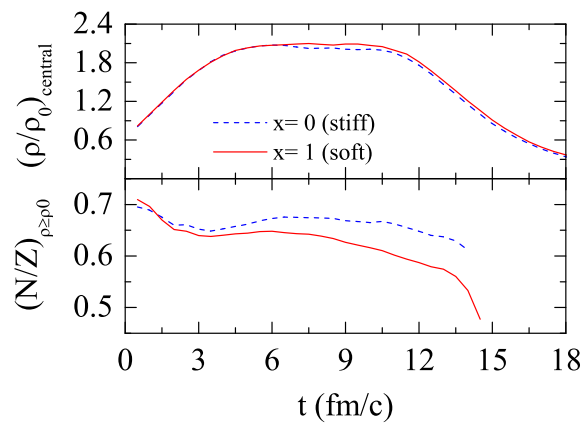}
\end{center}
\caption{(Color online) Central baryon density (upper window) and
isospin asymmetry N/Z of high density region (lower window) for
the reaction $^{22}$Si+$^{22}$Si at a beam energy of 400
MeV/nucleon and an impact parameter of 0 fm.} \label{hdnz}
\end{figure}
Fig.~\ref{hdnz} shows the central baryon density (upper window)
and the average $(N/Z)_{\rho\geq \rho_0}$ ratio (lower window) of
all regions with baryon densities higher than $\rho_0$. It is seen
that the maximum baryon density is about 2 times normal nuclear
matter density. Moreover, the compression is rather insensitive to
the symmetry energy because the latter is relatively small
compared to the EOS of symmetric matter around this density. The
high density phase lasts for about 11 fm/c (from 2 to 13 fm/c) for
this reaction. It is interesting to see that the isospin asymmetry
of the high density region is quite sensitive to the symmetry
energy. The soft symmetry energy ($x=1$) leads to a significantly
lower value of $(N/Z)_{\rho\geq \rho_0}$ than the stiff one ($x=
0$). This is consistent with the well-known isospin fractionation
phenomenon. Because of the $E_{sym}(\rho)\delta^2$ term in the EOS
of asymmetric nuclear matter, it is energetically more favorable
to have a larger isospin asymmetry $\delta$ in the high density
region with a softer symmetry energy functional $E_{sym}(\rho)$.
In the supranormal density region, as shown in Fig.\ 1 of
reference \cite{LYZ}, the symmetry energy changes from being soft
to stiff when the parameter $x$ varies from 1 to 0. Thus the value
of $(N/Z)_{\rho\ge \rho_0}$ becomes smaller as the parameter $x$
changes from 1 to 0. It is worth mentioning that the initial value
of the quantity $(N/Z)_{\rho\ge \rho_0}$ is about 0.7 which is
larger than the average n/p ratio of 0.57 of the reaction system.
This is because at lower density region of proton-rich nucleus
$^{22}_{14}$Si, there are more protons than neutrons as shown in
Fig.~\ref{dis}.

\begin{figure}[th]
\begin{center}
\includegraphics[width=0.45\textwidth]{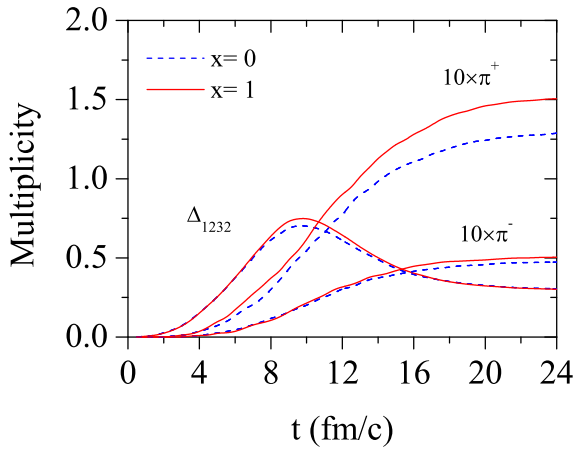}
\end{center}
\caption{(Color online) Evolution of $\pi^{-}$, $\pi ^{+}$ and
$\Delta(1232)$ multiplicities in the reaction $^{22}$Si+$^{22}$Si
at a beam energy of 400 MeV/nucleon and an impact parameter of 0
fm.} \label{multi}
\end{figure}
\begin{figure}[th]
\begin{center}
\includegraphics[width=0.45\textwidth]{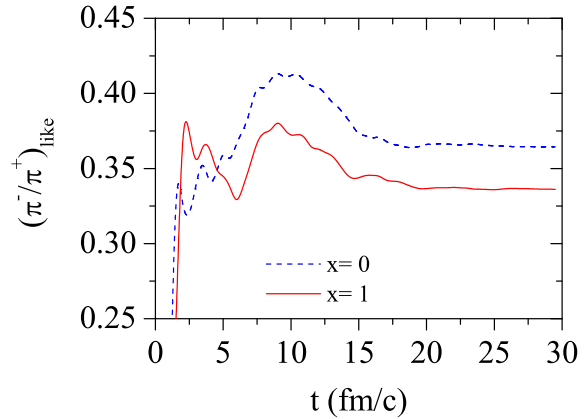}
\end{center}
\caption{(Color online) Evolution of the $\pi^-/\pi^+$ ratio in
the reaction $^{22}$Si+$^{22}$Si at a beam energy of 400
MeV/nucleon and an impact parameter of 0 fm.} \label{tratio}
\end{figure}

To understand the dynamics of pion production and its dependence
on the symmetry energy, we show in Fig.~\ref{multi} the
multiplicity of $\pi^+$, $\pi^-$ and $\Delta(1232)$ as a function
of time. The multiplicity of $\Delta(1232)$ resonances shown in
the figure includes all four charge states while in the model we
do treat and follow separately different charge states of the
$\Delta(1232)$. Because the value of $\pi^{-}/\pi^{+}$ becomes
stable after 30 fm/c (can be seen from Fig.~\ref{tratio}), here we
just give  evolutions of $\Delta(1232)$ and charged pions till 30
fm/c. At a beam energy of 400 MeV/nucleon which is just about 100
MeV above the pion production threshold in nucleon-nucleon
scatterings, almost all pions are produced through the decay of
$\Delta(1232)$ resonances. The contribution due to $N^*$
resonances is negligible. It is interesting to see that the
$\pi^{+}$ multiplicity depends more sensitively on the symmetry
energy. This is because the $\pi^{+}$ mesons are mostly produced
from proton-proton collisions, where asymmetry is always larger in
the reaction induced by the proton-rich nuclei
$^{22}$Si+$^{22}$Si. We can also see more $\pi^{+}$ than $\pi^{-}$
mesons are produced. Our finding that in proton-rich reactions
$\pi^{+}$ mesons are more sensitive to the symmetry energy than
$\pi^{-}$ contradicts the results of Ref. \cite{LYZ}.

To reduce the systematic errors in simulations, especially in
experimental analysis, one usually studies the $\pi^{-}/\pi ^{+}$
\cite{yong06,LYZ,Rei07} instead of $\pi^{-}$ or $\pi^{+}$ only.
Shown in Fig.~\ref{tratio} is effect of symmetry energy on the
$(\pi^-/\pi^+)_{like}$ as a function of time in the central
reaction $^{22}$Si+$^{22}$Si at a beam energy of $400$
MeV/nucleon. In the dynamics of pion resonance productions and
decays the $(\pi^-/\pi^+)_{like}$ reads \cite{LYZ}
\begin{equation}
(\pi^-/\pi^+)_{like}\equiv
\frac{\pi^-+\Delta^-+\frac{1}{3}\Delta^0}
{\pi^++\Delta^{++}+\frac{1}{3}\Delta^+}.
\end{equation}
This ratio naturally becomes $\pi^-/\pi^+$ ratio at the freeze-out
stage \cite{LYZ}. From Fig.~\ref{tratio} we can see that
sensitivity of $(\pi^-/\pi^+)_{like}$ to the effect of symmetry
energy is clearly shown after $t=10 fm/c$. With the stiff symmetry
($x= 0$) the value of $\pi^-/\pi^+$ is higher than that with the
soft symmetry ($x= 1$), this is understandable within the
statistical model for pion production \cite{LYZ}.

\begin{figure}[th]
\begin{center}
\includegraphics[width=0.45\textwidth]{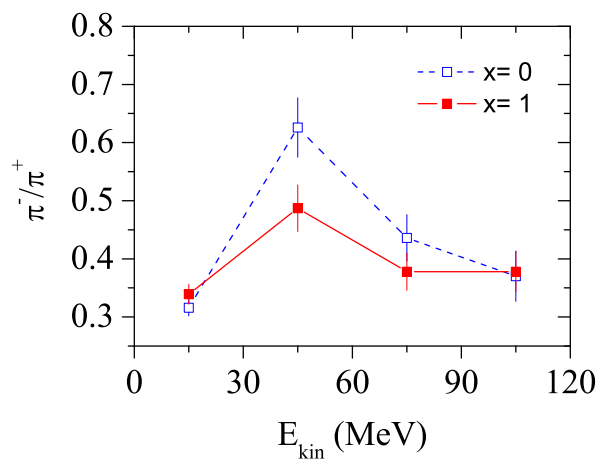}
\end{center}
\caption{(Color online) The $\pi^-/\pi^+$ ratio as a function of
pion kinetic energy in the reaction $^{22}$Si+$^{22}$Si at a beam
energy of 400 MeV/nucleon and an impact parameter of 0 fm.}
\label{eratio}
\end{figure}
\begin{figure}[th]
\begin{center}
\includegraphics[width=0.45\textwidth]{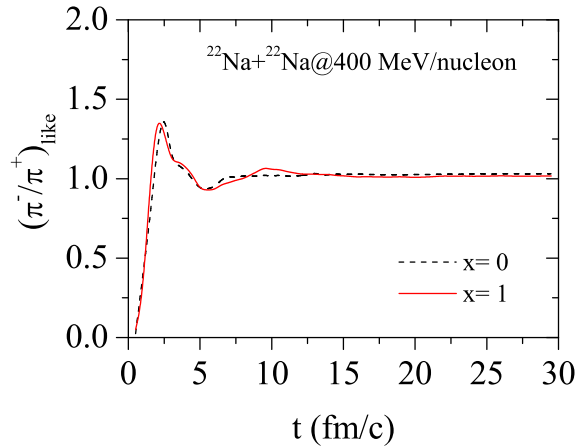}
\end{center}
\caption{(Color online) Evolution of the $\pi^-/\pi^+$ ratio in
the reaction $^{22}$Na+$^{22}$Na at a beam energy of 400
MeV/nucleon and an impact parameter of 0 fm.} \label{natratio}
\end{figure}

Shown in Fig.~\ref{eratio} is the differential $\pi^-/\pi^+$
ratios versus the kinetic energy. In the low energy ($E_{kin}\sim
45$ MeV) region, around the Coulomb peak \cite{yong06} the
$\pi^-/\pi^+$ ratio is clearly separable with the $x$ parameter
varying from 1 to 0. Sensitivity of $\pi^-/\pi^+$ ratio to the
symmetry energy around the Coulomb peak is about 20\%. In the
practical experiments, the proton-rich reaction
$^{22}$Si+$^{22}$Si may be difficult to carry out. The only
purpose of choosing this reaction is that we just want to show our
studies more clearly. In the practical experimental plan, a
reaction with proton-rich nucleus ($Z>N$) and stable nucleus
($Z\sim N$) (such as $^{22}$Si+$^{40}$Ca) is feasible.
\begin{figure}[th]
\begin{center}
\includegraphics[width=0.45\textwidth]{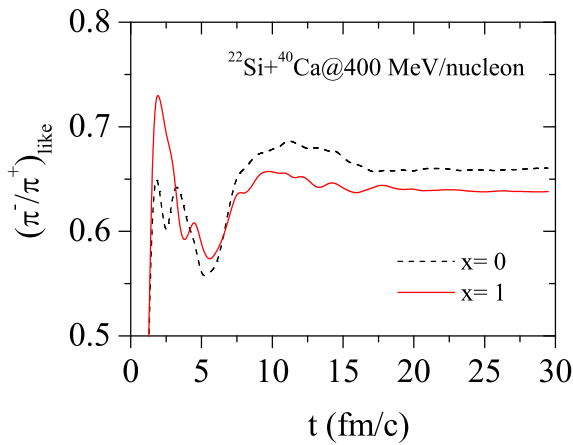}
\end{center}
\caption{(Color online) Evolution of the $\pi^-/\pi^+$ ratio in
the reaction $^{22}$Si+$^{40}$Ca at a beam energy of 400
MeV/nucleon and an impact parameter of 0 fm.} \label{catratio}
\end{figure}
Shown in Fig.~\ref{catratio} is the effect of symmetry energy on
the $(\pi^-/\pi^+)_{like}$ as a function of time in the central
reaction $^{22}$Si+$^{40}$Ca at a beam energy of $400$
MeV/nucleon. Effect of nuclear symmetry energy on $\pi^-/\pi^+$ is
clearly shown. Comparing Fig.~\ref{catratio} with
Fig.~\ref{tratio}, we can see that the effect of symmetry energy
is smaller for $^{22}$Si+$^{40}$Ca reaction than for
$^{22}$Si+$^{22}$Si. This is due to a smaller asymmetry (absolute)
value of $^{22}$Si+$^{40}$Ca. As a comparison, we also provide
Fig.~\ref{natratio}, the case of \emph{zero} asymmetry reaction
system $^{22}$Na+$^{22}$Na ($^{22}$Na's half-life is 2.6 years).
As one expects, there is no symmetry energy effect on charged pion
ratio in the reaction $^{22}$Na+$^{22}$Na. The $\pi^-/\pi^+$ ratio
is approximately $1$, roughly equal to $(5N^{2} + NZ)/(5Z^{2} +
NZ) \sim (N/Z)^{2} = (44/44)^{2} = 1$ in central heavy-ion
reactions \cite{yong06}, with N and Z being the total neutron and
proton numbers in the participant region.


In conclusion, based on an isospin dependent transport model IBUU,
effects of high density behavior of nuclear symmetry energy on
$\pi^{-}/\pi ^{+}$ in the proton-rich reactions
$^{22}$Si+$^{22}$Si, $^{22}$Si+$^{40}$Ca at a beam energy of $400$
MeV/nucleon are studied. It is found that the symmetry energy
evidently affects $\pi^{+}$ production and the value of
$\pi^{-}/\pi^{+}$. As one expected, comparing with neutron-rich
reactions, effect of symmetry energy in the proton-rich reaction
shows contrary behavior. Studying proton-rich rations can not only
help us to probe the symmetry energy, but also check the theories
about nuclear matter.


The work is supported by the National Natural Science Foundation
of China (10875151, 10740420550), the Knowledge Innovation Project
(KJCX2-EW-N01) of Chinese Academy of Sciences, the Major State
Basic Research Developing Program of China under No. 2007CB815004,
and the CAS/SAFEA International Partnership Program for Creative
Research Teams (CXTD-J2005-1).


\begin{thebibliography}{00}

\bibitem{xiao09}Z.G. Xiao, B.A. Li, L.W. Chen, G.C. Yong, M. Zhang, Phys. Rev.
Lett. {\bf 102} 062502 (2009).

\bibitem{ditoro1}M. Di Toro, V. Baran, M. Colonna, V. Greco, J. Phys. G: Nucl.
Phys. {\bf 37}, 083101 (2010).

\bibitem{xu09}J. Xu, C.M. Ko, Y. Oh, Phys. Rev. {\bf C81}, 024901 (2010).

\bibitem{Rei07}W. Reisdorf, M. Stockmeier, A. Andronic, M.L. Benabderrahmane,
O.N. Hartmann, N. Herrmann, K.D. Hildenbrand, Y.J. Kima, et al.,
Nucl. Phys. {\bf A781}, 459 (2007).

\bibitem{yong06}G.C. Yong, B.A. Li, L.W. Chen, W. Zuo, Phys. Rev. {\bf C73}, 034603 (2006).

\bibitem{LiBA02}B.A. Li, Phys. Rev. Lett. {\bf 88}, 192701 (2002).

\bibitem{Bro00}B.A. Brown, Phys. Rev. Lett. {\bf 85}, 5296 (2000).

\bibitem{Dan02a}P. Danielewicz, R. Lacey, W.G. Lynch, Science {\bf 298}, 1592 (2002).

\bibitem{Bar05}V. Baran, M. Colonna, V. Greco, M. Di Toro, Phys. Rep. {\bf 410}, 335 (2005).

\bibitem{LCK08}B.A. Li, L.W. Chen and C.M. Ko, Phys. Rep. {\bf 464}, 113 (2008).

\bibitem{Sum94}K. Sumiyoshi and H. Toki, Astrophys. J. {\bf 422}, 700 (1994).

\bibitem{Lat04}J.M. Lattimer, M. Prakash, Science {\bf 304}, 536 (2004).

\bibitem{Ste05a}A.W. Steiner, M. Prakash, J.M. Lattimer, P.J. Ellis, Phys. Rep. {\bf 411}, 325 (2005).

\bibitem{Kut94}M. Kutschera, Phys. Lett. {\bf B340}, 1 (1994).

\bibitem{Kub99}S. Kubis and M. Kutschera, Acta Phys. Pol. {\bf B30},
2747 (1999); Nucl. Phys. {\bf A720}, 189 (2003).

\bibitem{Che07}L.W. Chen, C.M. Ko, B.A. Li, Phys. Rev. {\bf C76}, 054316 (2007).

\bibitem{LiZH06}Z.H. Li, U. Lombardo, H.J. Schulze, W. Zuo, L.W. Chen, and H.R. Ma, Phys. Rev. {\bf C74}, 047304 (2006).

\bibitem{Pan72}V.R. Pandharipande, V.K. Garde, Phys. Lett. {\bf B39}, 608 (1972).

\bibitem{Fri81}B. Friedman, V.R. Pandharipande, Nucl. Phys. {\bf A361}, 502 (1981).

\bibitem{Wir88a}R.B. Wiringa, V. Fiks, Phys. Rev. {\bf C38}, 1010 (1988).

\bibitem{Kra06}P. Krastev and F. Sammarruca, Phys. Rev. {\bf C74}, 025808 (2006).

\bibitem{Szm06}A. Szmaglinski, W. Wojcik, M. Kutschera, Acta Phys. Polon. {\bf B37}, 227 (2006).

\bibitem{Cha97}E. Chabanat, P. Bonche, P. Haensel, J. Meyer, R. Schaeffer, Nucl. Phys. {\bf A627} (1997) 710; {\it ibid}, 635 (1998) 231.

\bibitem{Sto03}J.R. Stone, J.C. Miller, R. Koncewicz, P. D. Stevenson, M. R. Strayer, Phys. Rev. {\bf C68}, 034324 (2003).

\bibitem{Che05b}L.W. Chen, C.M. Ko and B.A. Li, Phys. Rev. {\bf C72}, 064309 (2005).

\bibitem{Dec80}J. Decharge and D. Gogny, Phys. Rev. {\bf C21}, 1568 (1980).

\bibitem{MS}W.D. Myers and W.J. Swiatecki, Acta Phys. Pol. {\bf B26},111 (1995).

\bibitem{Kho96}D.T. Khoa, W. von Oertzen, A.A. Ogloblin, Nucl. Phys. {\bf A602}, 98 (1996).

\bibitem{Bas07}D.N. Basu, T. Mukhopadhyay, Acta Phys. Polon. {\bf B38}, 169 (2007).

\bibitem{Ban00}S. Banik and D. Bandyopadhyay, J. Phys. G {\bf 26}, 1495 (2000).

\bibitem{tsang09}M.B. Tsang, Y.X Zhang, P. Danielewicz, M. Famiano, Z.X. Li, W.G. Lynch, and A.W. Steiner, Phys. Rev. Lett. {\bf 102}, 122701 (2009).

\bibitem{shetty07}D.V. Shetty, S. J. Yennello, and G. A. Souliotis, Phys. Rev. {\bf C76}, 024606 (2007).

\bibitem{fami06}M.A. Famiano, T. Liu, W.G. Lynch, M. Mocko, A.M. Rogers, M.B. Tsang, M.S. Wallace, R.J. Charity, S. Komarov, D.G. Sarantites, L.G. Sobotka, and G. Verde, Phys. Rev. Lett. {\bf 97}, 052701 (2006).

\bibitem{tsang04}M.B. Tsang, T.X. Liu, L. Shi, P. Danielewicz, C.K. Gelbke, X.D. Liu, W.G. Lynch, W.P. Tan, G. Verde, A. Wagner, and H.S. Xu, Phys. Rev. Lett. {\bf 92}, 062701 (2004).

\bibitem{chen05}L.W. Chen, C.M. Ko and B.A. Li, Phys. Rev. Lett. {\bf 94}, 032701 (2005).

\bibitem{Sto86}R. Stock, Phys. Rep., {\bf 135}, 259 (1986).

\bibitem{Ber80}G.F. Bertsch, Nature {\bf 283}, 280 (1980);
A. Bonasera and G.F. Bertsch, Phys. Let. B195 (1987) 521.

\bibitem{Gai04}T. Gaitanos, M. Di Toro, S. Typel, V. Baran, C. Fuchs, V. Greco, H.H. Wolter, Nucl. Phys. {\bf A732}, 24 (2004).

\bibitem{LiQF05b}Q.F. Li, Z.X. Li, S. Soff, M. Bleicher, and Horst St\"{o}cker, Phys. Rev. {\bf C72}, 034613 (2005).

\bibitem{Das03}C. B. Das, S. Das Gupta, C. Gale, and B.A. Li, Phys. Rev. {\bf C67}, 034611 (2003).

\bibitem{IBUU04}B.A. Li, C.B. Das, S. Das Gupta, C. Gale, Nucl. Phys. {\bf A735}, 563 (2004); Phys. Rev. {\bf C69}, 064602
(2004).

\bibitem{neg}J.W. Negele and K. Yazaki, Phys. Rev. Lett. {\bf 47}, 71
(1981).

\bibitem{pan}V.R. Pandharipande and S.C. Pieper, Phys. Rev. {\bf C45}, 791
(1991).

\bibitem{gale}D. Persram and C. Gale, Phys. Rev. {\bf C65}, 064611 (2002).

\bibitem{factor}B.A. Li and L.W. Chen, Phys. Rev. {\bf C72}, 064611 (2005).

\bibitem{ditoro2}V. Prassa, G. Ferini, T. Gaitanos, H.H. Wolter, G.A. Lalazissis, M. Di Toro, Nucl. Phys. {\bf A789}, 311 (2007).

\bibitem{Lar01}A.B. Larionov, W. Cassing, S. Leupold, U. Mosel, Nucl. Phys. {\bf A696}, 747 (2001).

\bibitem{Lar03}A.B. Larionov, U. Mosel, Nucl. Phys. {\bf A728}, 135 (2003).

\bibitem{Ber88}G.F. Bertsch, G.E. Brown, V. Koch, B.A. Li, Nucl. Phys. {\bf A490}, 745 (1988).

\bibitem{Mao97}G.J. Mao, Z.X. Li, Y.Z. Zhuo and E.G. Zhao, Phys. Rev. {\bf C55}, 792 (1997).

\bibitem{Friedrich86}J. Friedrich and P.G. Reinhard, Phys. Rev. {\bf C33}, 335 (1986).

\bibitem{yong1103}G.C. Yong, Y. Gao, W. Zuo, X.C. Zhang, arXiv: 1104.0103 (2011).

\bibitem{Ch97}S. Chakrabarty, D. Bandyopadhyay and S. Pal, Phys. Rev. Lett. {\bf 78},
2898 (1997).

\bibitem{LYZ}B.A. Li, G.C. Yong and W. Zuo, Phys. Rev. {\bf C71}, 014608 (2005).


\end{thebibliography}
\end{document}